# Why the future cannot be open in the quantum world

Kunihisa Morita[1]

**Abstract**

In this study, I argued that the future is not open if quantum mechanics (QM) is complete. An open future means that the value observed when measuring a physical quantity in the future is not determined at present. At first glance, QM seems to support the open future thesis because it cannot always predict the measured value with certainty. However, many interpretations regarding QM are deterministic, suggesting that the quantum mechanical world can also be deterministic. I argued that, although QM cannot predict the future with certainty, the quantum mechanical world *must be* deterministic, and the value observed by the observer is determined. I examined the following two cases: (1) the wave function completely describes the physical state and (2) the wave function does not describe the physical state. Then, I argued that the future cannot be open in either case when QM is complete.

## 1. Introduction

Generally, an open future means that the events of the future are not determined at present, suggesting that propositions stating the future events do not have definite truth-values at present, which will be further discussed later. In this paper, I argue that if

---

[1] Graduate School of Human Sciences, Osaka University, JAPAN.

Email: kunihisa@hus.osaka-u.ac.jp

quantum mechanics (QM) is a complete theory, then the future is not open, contrary to appearances. When theory $T$ is complete, any information other than which theory $T$ requires is unnecessary to explain the experimental data. That is, there are no "hidden variables." Therefore, since QM does not predict the measurement values of physical quantities with certainty except in specific cases, the world is essentially probabilistic if QM is complete unless some interpretations are accepted, such as Bohmian mechanics (Bohm 1952a; 1952b) and the many-worlds interpretation.

Accordingly, at first, QM strongly supports the open future thesis. However, there is a possibility that a physical quantity (observable), $Q$, has one definite value before measurement; however, in principle, it is impossible to predict which value of $Q$ the observer will find through measurement. There are many interpretations of QM insisting that physical quantities that are not yet measured have definite values, such as modal interpretations. However, as discussed below (§3), those interpretations do not necessarily exclude the open future thesis initially and their proponents do not intend to reject the open future thesis; rather, they would support realism. Therefore, I argue in this paper that if QM is complete and there is no non-physical process, QM cannot be interpreted as a theory that supports the open future thesis. Non-physical process refers to a process that cannot be reduced into a physical one. In other words, accepting the existence of non-physical process implies the acceptance of mind–body dualism that is not accepted by modern physicists.

For the rest of this section, I clarify the concept of the open future. $F(x)$, $O(x, y)$, and $\Box$ are as follows:

$F(x)$: "in $x$ time units it will be the case that …"

$O(y, z)$: "an observer observes one definite value, $y$, by the measurement of a physical quantity, $z$"

□: "it is now necessarily the case that …"

Therefore, the open future thesis is defined as

$$\neg\Box F(x)O(y, z) \wedge \neg\Box F(x) O(\neg y, z). \quad [\text{OP}]$$

The open future thesis means both that it is not necessarily the case now that the observer will obtain one definite value $y$ by measuring observable $z$ in time $x$ in the future, and that it is not necessarily the case now that the observer will not obtain one definite value $y$ by measuring $z$ in time $x$ in the future.

$Q$ possesses one definite value $a$, which is expressed as[1]

$$\Box O(a, Q). \quad [\text{D}]$$

It is now necessarily the case that the observer obtains definite value $a$ by the measurement of $Q$ [2]. Here, it is assumed that the observer always obtains one definite value of $Q$ by the measurement:

$$\exists x, O(x, y). \quad [\text{AD}]$$

For example, interpretations such as the many-worlds interpretation and many-minds interpretation assume that the observer cannot arrive at one definite value of $Q$ through measurement. The observer believes that they are observing one definite value, but that is an illusion (thus, condition [AD] is false). However, such cases are not open future situations, even if [OP] is satisfied, because the undetermined future will become a *determined* present and past according to the open future thesis. Therefore, both [OP] and [AD] are necessary for the open future thesis.

This paper consists of two parts. First, I argue that if the wave function completely describes the physical state of the system, then we have to accept that non-physical processes causally influence the physical processes (thus, the interpretation that the wave function completely describes the physical state is unacceptable); this interpretation can support the open future thesis (§2). Second, I argue that those interpretations of QM in which the wave function does not describe the physical state cannot support the open future thesis (§3).

**2. Wave function completely describes the physical state**

In this section, I assume that the wave function completely describes the *physical* state of a physical system. The implications of this are as follows. Suppose $\varphi(t)$ is a wave function of system $S$ at a certain point in time $t$. Suppose $t_1$ and $t_2$ are the points in time and $t_2$ is further in the future than $t_1$. Possible values (eigenvalues) of physical quantity $Q$ are $q_i$ ($i = 1, 2, …, N$), and their corresponding eigenstates are $|q_i>$. Now, suppose the following holds:

$$\varphi(t_2) = a_1(t_2)|q_1> + … + a_N(t_2)|q_N>, \qquad (2\text{-}1)$$

where $a_i$ are normalized constants. For $\varphi(t_2)$ to completely describe the physical state of $S$ means that $Q$ does not possess any definite value at $t_2$. Therefore, when $t_1$ is the present, then

$$\neg\Box F(t_2 - t_1)O(q_1, Q) \land \ldots \land \neg\Box F(t_2 - t_1)O(q_k, Q) \land \ldots \land \neg\Box F(t_2 - t_1)O(q_N, Q). \quad (2\text{-}2)$$

However, the observer observes one definite value (e.g., $q_1$) when $t_2$ becomes the present and $Q$ is measured. Thus, $Q$ possesses the definite value ($q_1$) at $t_2$, which is expressed as

$$\Box O(q_1, Q). \quad (2\text{-}3)$$

Since the wave function completely describes the physical state of $S$, the wave function changes to

$$\varphi(t_2) = |q_1\rangle. \quad (2\text{-}4)$$

In contrast, (2-3) is true if (2-4) is true, and (2-1) is true if (2-2) is true. These relations are referred to as the "eigenstate-eigenvalue link." In other words, the eigenstate-eigenvalue link must work for the wave function to completely describe the physical state.

The interpretation I have previously discussed is called the orthodox interpretation or standard interpretation. The features of this interpretation are as follows.

(a) QM is complete;

(b) The Schrödinger equation, which is the fundamental equation of QM, completely describes the temporal behavior of the wave function, except in the measurement process;

(c) The observer obtains one definite value by measurement;

(d) Physical quantity $Q$ possesses a definite value if and only if the wave function is an eigenfunction of $Q$ (eigenstate-eigenvalue link);

(e) The square of probability to observe definite value $q_i$ is $a_i^* a_i$ (Born Rule).

For (2-2) and (2-3), the orthodox interpretation satisfies the conditions for the open future thesis, [OP] and [AD].

However, the transition from (2-1) to (2-4) is mysterious. This instantaneous and discontinuous change from (2-1) into (2-4) is called the "collapse of the wave function." The process of the collapse of the wave function cannot be described by the Schrödinger equation because this process is discontinuous, and the Schrödinger equation is a continuous differential equation. Therefore, Wigner (1961) argued the orthodox interpretation of QM, which is characterized by (a)–(e), implying mind–body dualism.[3]

If there is no non-physical process, the measurement process is also a physical process. If the measurement process is a physical process, it must be described by the Schrödinger equation because there is no reason to distinguish between measurement and physical processes. Nevertheless, the measurement process cannot be considered a physical process since the Schrödinger equation cannot describe the measurement process as discussed above. Therefore, the orthodox interpretation implies that non-physical processes causally influence physical ones (e.g., consciousness of observer causes the collapse of the wave function).

Note that I do not insist that this discontinuous process cannot be described mathematically. This process can be described assuming *additional* assumption (i.e., the projection postulate). However, the point is that there is no reason to apply this postulate only to the measurement process because the measurement process cannot be distinguished from other physical processes if both are physical processes.

However, proponents of the view that the wave function can completely describe the physical state might insist that the reason the Schrödinger equation cannot describe the measurement process (there seems to be the collapse of the wave function) is that the Schrödinger equation is applied only to a closed system, whereas the system in question *S* is not a closed system owing to its interaction with the measurement apparatus and the observer. Certain physicists and philosophers of physics try to explain the collapse of the wave function by arguing that the interaction between the system and measurement apparatus, including the observer, causes the collapse of the wave function (Myrvold 2018, §2.3.2). However, supposing that the Schrödinger equation could completely describe the behavior of the wave function of the whole system (*S*) and its environment system, including the measurement apparatus and observers, and that the Schrödinger equation could deterministically predict the state of this whole system, this would also contradict the assumption that the future is open ([OP] is not satisfied). Note that I do not object this view (the Schrödinger equation considering the whole system can describe the measurement process). I insist that if this view is correct, it indicates that the quantum mechanical world is not the open future world.

Therefore, if QM is complete and the future is open, the measurement process that changes the indefinite value of $Q$ to a definite value cannot be a physical process. There is a possibility that there are some non-physical processes in the world (the mind–body dualism is true). However, physics is not closed if such non-physical processes can causally influence physical processes. For physics not to be closed implies, that physics cannot explain all physical phenomena, a hypothesis that is difficult to accept in general.

## 3. Wave function does not completely describe the physical state

As discussed in §2, if the wave function describes the physical state and the future is open, then assumption that the non-physical processes influence physical processes must be accepted. However, one does not need to accept the assumption that the wave function completely describes the *physical* state, even if the completeness of QM is accepted.

Furthermore, van Fraassen (1991) suggests that there are two states of systems, that is, the dynamic and value states. The dynamic state determines the physical properties (values) the system may possess and those the system may have at later times. Thus, the wave function represents the dynamic state (the wave function describes the state of our *knowledge* of a physical system). On the other hand, the value state represents what is the actual case (i.e., all physical properties of the system that are sharply defined at the instant in question) (Lombardi & Dieks 2017). This idea is common to "modal interpretations." An essential feature of this interpretation is that physical quantity $Q$ can possess a definite value even if the dynamical state is not an eigenstate of $Q$, indicating that the wave function is not an eigenfunction of $Q$.

Therefore, the eigenstate-eigenvalue link is violated. This possibly suggests that [OP] is not satisfied even if the wave function at $t$ (in the future) is not an eigenstate of $Q$ that will be measured at $t$.

However, there seems to be a loophole to save the open future thesis, contrary to appearances. Suppose $t_1$ is the present point in time and $t_2$ and $t_3$ are points in time in the future ($t_1 < t_2 < t_3$). In addition, suppose electron $e_1$ interacts with electron $e_2$ at $t_2$, they spatially separate soon after $t_2$, and no external forces act on $e_1$ and $e_2$ between $t_2$ and $t_3$. After the interaction, the spin state (dynamic state) of the system consisting of $e_1$ and $e_2$ is $|+1/2>_I |-1/2>_{II} - |-1/2>_I |+1/2>_{II}$ (where, $|>_I$ and $|>_{II}$ represent the states of $e_1$ and $e_2$, respectively, the normalized constants are ignored, and the unit is $\hbar$). $e_1$ and $e_1$ are entangled. However, according to the modal interpretation, the value state of $e_1$ must be either $|+1/2>_\sigma$ or $|-1/2>_\sigma$ ($\sigma = x, y, z$), and the $e_1$ and $e_2$ systems are separated.

Therefore, when $t_2$ is the present point in time, and the observer measures the $x$-spin of $e_1$ at $t_3$, the following holds:

$$\Box F(t_3 - t_2)O(+1/2, x\text{-spin}) \lor \Box F(t_3 - t_2)O(-1/2, x\text{-spin}) \qquad (3\text{-}1)$$

Therefore, at $t_2$, the future is not open with respect to the $x$-spin value of $e_1$. However, there is no guarantee that (3-2) is also true when $t_1$ is present.

$$\Box F(t_3 - t_1)O(+1/2, x\text{-spin}) \lor \Box F(t_3 - t_1)O(-1/2, x\text{-spin}), \qquad (3\text{-}2)$$

implying that what will happen (value of the *x*-spin the observer will observe) at $t_3$ might not be determined at $t_1$ (before the interaction of $e_1$ and $e_2$). Accordingly, there is still a possibility that the future is open at $t_1$.

Nevertheless, I argue that the future cannot be open according to the modal interpretation. Generally, the Kochen–Specker (K–S) theorem states that all physical quantities cannot have definite values simultaneously if QM is complete (Kochen & Specker 1967). Thus, (3-1) and the following propositions, (3-3) and (3-4), cannot simultaneously be true when $t_2$ is the present.

$$\Box F(t_3 - t_2)O(+1/2, \textit{y-spin}) \lor \Box F(t_3 - t_2)O(-1/2, \textit{y-spin}). \qquad (3\text{-}3)$$

$$\Box F(t_3 - t_2)O(+1/2, \textit{z-spin}) \lor \Box F(t_3 - t_2)O(-1/2, \textit{z-spin}). \qquad (3\text{-}4)$$

This indicates that the physical quantities that can possess definite values are restricted by a certain rule. Although different versions of modal interpretation have different rules, the physical quantity that will be measured must possess a definite value; otherwise, the same problem discussed in §2 arises.

Therefore, if the observer aims, at $t_2$, to measure the *x*-spin at $t_3$, (3-1) is true while (3-3), (3-4), or both (3-3) and (3-4) are not true. According to the K–S theorem, these three observables cannot possess definite values simultaneously. On the contrary, if the observer aims, at $t_2$, to measure the *y*-spin at $t_3$, (3-3) is true while (3-1), (3-4), or both (3-1) and (3-4) are not true. However, suppose the observer changes his/her mind at $t_4$ ($t_2 < t_4 < t_3$) and instead decides to measure the *y*-spin at $t_3$, although the observer aims at $t_2$ to measure the *x*-spin at $t_3$, and the observer again changes his/her mind at $t_5$ ($t_4 < t_5 < t_3$) to measure the *z*-spin at $t_3$, then the truth value of at least one among (3-1), (3-3), and (3-4) changes between $t_2$ and $t_3$. This is unacceptable since no external force on $e_1$ (and $e_2$) between $t_2$ and $t_3$ is assumed. In addition, this seems to violate the concept of "necessary truth." To emphasize this absurdity, "EPR situation" is considered, which was originally suggested by Einstein et al. (1945) and revised by Bohm (1989, pp. 611ff.).

In addition, consider the state of $e_2$. Suppose the total spin of $e_1$ and $e_2$ is 0 at $t_2$. $t_4$ (and thus $t_3$) is long enough after $t_2$, and $e_2$ is a few light years away from $e_1$ and the measurement apparatus on the earth. The observer set the measurement apparatus for measuring the *x*-spin of $e_1$ at $t_1$. Thus, both the *x*-spins of $e_1$ and $e_2$ possess definite values at $t_2$ according to the modal interpretation, while either the *y*-spin of $e_1$ and $e_2$ (or both) does not possess a definite value at $t_2$ (for the convenience of discussion, we assume that *x*-spin and *y*-spin cannot simultaneously possess definite values because of the K–S theorem). Thus far, this is not strange at all. However, the observer changes the set of the measurement apparatus such that it is appropriate for measuring the *y*-spin at $t_4$. Then, the *y*-spin of $e_2$ begins to possess a definite value at $t_4$ if it does not possess one, despite the fact that $e_2$ is far from both $e_1$ and the measurement apparatus (a case where the observable that does not possess one definite value is the *z*-spin, and the same line of argument is applied).

Note that there is no quantum non-local correlation in this situation because the value states of the *x*(*y*)-spins of $e_1$ and $e_2$ are either $|+1/2\rangle_{x(y)}$ or $|-1/2\rangle_{x(y)}$ (i.e., they are not the quantum entangled states $|+1/2\rangle_I |-1/2\rangle_{II} - |-1/2\rangle_I |+1/2\rangle_{II}$). Therefore, there must be some superluminal *mechanical* interaction between $e_1$ and $e_2$ (or between the

measurement apparatus and $e_2$) because these states are physical states. However, this is cannot be realized according to the relativity theory.[4]

To avoid such an absurd conclusion, the exact physical quantity the observer will measure at $t_3$ must be determined before $t_2$. Although the observer can change his/her mind between $t_1$ and $t_3$, the final state of his/her mind at $t_3$ is determined before $t_2$. In the above discussion leading to an absurd conclusion, we have presupposed the open future thesis.

$$\neg\Box F(x)M(y) \land \neg\Box F(x)\neg M(y), \qquad (3\text{-}5)$$

where $M(y)$ is the propositional function that the observer measures physical quantity $y$, indicating that the observable the observer will measure in the future is not determined now.

In our current problem, the $x$-spin of $e_1$ might not possess one definite value before $t_2$. Specifically, (3-2) is possibly false; thus, the open future thesis might be correct even under the modal interpretations. The physical quantity that the observer will measure is determined before $t_2$. Nevertheless, (3-2) does not seem to be true. However, what is the significance of measuring the physical quantity by the observer at $t_3$ being determined before $t_2$? This indicates that the state of the mind of the observer at $t_3$ has been determined before $t_2$.[5] The brain consists of many quantum particles, and these particles must interact with each other many times before $t_3$. As discussed, the physical quantity the observer will measure is determined. It follows that the definite values possessed by the physical quantities (elements of the brain of the observer) are also determined before the interactions of these particles in the brain of the observer.

Therefore, the definite value that will be possessed by the *x*-spin of $e_1$ must also be determined before $t_2$ because there is no fundamental difference between electron system $e_1$ and the brain system of the observer. Therefore, (3-2) is true at $t_1$. In conclusion, the open future thesis is false under the modal interpretation.

However, while (3-2) is true, either (3-3) or (3-4) is false. This satisfies the condition for the open future [OP]. Nevertheless, remember that [AD] should be satisfied for the open future thesis. As previously discussed, the *y*-spin and *z*-spin of $e_1$ is not measured at $t_3$, and they cannot satisfy [AD] at $t_3$. They can be measured after $t_3$, and the observer observes a definite value of the *y*-spin or *z*-spin at $t_3$. However, in this case, one of the following is true.

$$\Box F(t_6 - t_2)O(+1/2, y\text{-spin}) \vee \Box F(t_6 - t_2)O(-1/2, y\text{-spin}), \quad (3\text{-}6)$$

$$\Box F(t_6 - t_2)O(+1/2, z\text{-spin}) \vee \Box F(t_6 - t_2)O(-1/2, z\text{-spin}), \quad (3\text{-}7)$$

where $t_6 > t_3$. As discussed above, whether (3-6) or (3-7) is true is determined before $t_2$.

Finally, there is another interpretation assuming that the wave function does not completely describe the physical state although accepting the completeness of QM. This interpretation is referred to as QBism (Fuchs 2010). QBism claims that the wave function describes the state of our knowledge, and it regards the actual physical state as a type of black box such that one can avoid committing to the unobserved state. QBism considers QM as a tool to explain experimental data; it suggests that QM does not need to offer any description of the unobserved state. Therefore, QBism can yield no metaphysical claims about future quantum states. Accordingly, I do not consider QBism in this paper.

**4. Summary**

If QM is complete, then the future must be open, or at least it might be open at first glance because measurement values are principally unpredictable according to QM. However, as I have argued in this paper, the future cannot be open if the QM is complete. An open future means both that it is not necessarily the case now that the observer will obtain a particular definite measurement value ($a$) and that it is not necessarily the case now that the observer will not obtain $a$ when the observer measures physical quantity $Q$ in the future.

      First, a case where the wave function describes the *physical* state is examined. Because in most cases the wave function of $Q$ at $t$ in the future is not an eigenfunction, $Q$ is considered to possess no definite value at $t$; thus, the future is open. However, the observer obtains one definite value of $Q$ when $t$ becomes the present point in time and measures $Q$. Therefore, this measurement process is discontinuous, and the Schrödinger equation cannot describe this process. If this measurement process is a physical one, it must be described by the Schrödinger equation because there is no fundamental difference between the measurement process and other physical processes that can be described by the Schrödinger equation. Therefore, it does not matter whether the discontinuous process can be described mathematically or not, the problem is whether this process can be described by the Schrödinger equation. Accordingly, if the future is open, non-physical processes (e.g., consciousness of the observer that cannot be reduced into a physical process) can influence the physical ones (namely, we must accept the

mind–body dualism); however, this is unacceptable. Nevertheless, there is a possibility that the measurement process can be described by the Schrödinger equation when considering the observer, measurement apparatus, and system in question (thus, we do not need to accept mind–body dualism). If this is the case, then the world is deterministic; thus, the future is not open. In conclusion, if QM is complete, the wave function describes the physical state and non-physical processes cannot influence the physical ones, indicating that the future cannot be open.

Second, I examined a case where the wave function does not completely describe the physical state (the wave function describes the state of our knowledge). Because this case does not assume the eigenstate-eigenvalue link, $Q$ can possess one definite value even when the wave function is not an eigenfunction of $Q$. However, not all physical quantities can possess definite values simultaneously because of the Kochen–Specker theorem. Nevertheless, physical quantities that will be measured possess definite values to avoid the discontinuous collapse of the wave function. This type of interpretation is called the modal interpretation. Although the modal interpretation seems to exclude the open future thesis, it is possible that the definite value of physical quantity $Q$ of system $S$ is not determined before system $S$ finally interacts with the external system. However, I argue that the physical quantity the observer will measure must be determined before the final interaction of $S$ with the external system. This conclusion shows us that the definite value of the physical quantity must be determined before the interaction. In conclusion, the future cannot be open if QM is complete.

**Acknowledgements**

*This paper is published in Principia 26(3)* <https://doi.org/10.5007/1808-1711.2022.e84794>

*Please cite the published version.*

I would like to thank two anonymous reviewers for their helpful comments. I also would like to thank Editage (www.editage.com) for English language editing.

**Notes**

---

[1] There are controversies regarding how to interpret the state of quantum mechanical indeterminacy, or the superpositional state (Darby 2010; Skow 2010; Torza 2017; Calosi & Wilson 2019). There arises the question of which is the adequate account: the supervaluationism account or the determinable-based account. However, definition [D] does not support either side of this debate, because it does not refer to state itself (rather, it only refers to *measurement* value).

[2] Therefore, rejecting an open future thesis does not mean accepting causal determinism. See (Morita 2020, pp. 57–58).

[3] Recently, Barrett (2006) and Morita (2020) also argued that the orthodox interpretation implies mind–body dualism.

[4] Although this interaction cannot transmit any information with superluminal speed, this clearly changes the *physical* state (the value state) of $e_2$ by a distant cause (the measurement apparatus). Note again that this is not a non-local correlation, but a mechanical interaction.

[5] One might object that the physical quantity that the observer will measure is determined at $t_2$ when $e_1$ and $e_2$ are intact. However, the observer can be far from $e_1$, $e_2$, and the measurement apparatus before and at $t_2$; thus, if the interaction of $e_1$ and $e_2$ influences the observer's brain, this can follow from the absurdity as discussed.